\documentstyle[amssymb,preprint,aps]{revtex}

\newtheorem{theorem}{Theorem}

\newtheorem{definition}[theorem]{Definition}

\begin{document}
\title{Harmonic maps and isometric embeddings of the spacetime }
\author{S. Chervon$^{a}$ , F. Dahia$^{b}$ and C. Romero$^{a}$}
\address{$^{a}$Departamento de F\'{i}sica, Universidade Federal da Para\'{i}ba, C.\\
Postal 5008, \\
58051-970 Jo\~{a}o Pessoa, Pb, Brazil\\
$^{b}$Departamento de F\'{i}sica, Universidade Federal de Campina Grande,\\
58109-970, Campina Grande, Pb, Brazil\\
E-mail: chervon@fisica.ufpb.br, fdahia@df.ufpb.br and cromero@fisica.ufpb.br}
\maketitle

\begin{abstract}
We investigate harmonic maps in the context of isometric embeddings when the
target space is Ricci-flat and has codimension one. With the help of the
Campbell-Magaard theorem we show that any $n$-dimensional ($n\geqslant 3$)
Lorentzian manifold can be isometrically and harmonically embedded in a
(n+1)-dimensional semi-Riemannian Ricci-flat space. We then extend our
analysis to the case when the target space is an Einstein space. Finally, as
an example, we work out the harmonic and isometric embedding of a
Friedmann-Robertson-Walker spacetime in a five-dimensional Ricci-flat space
and proceed to obtain a general scheme to minimally embed any vacuum
solution of general relativity in Ricci-flat spaces with codimension one.
\end{abstract}

\section{Introduction\protect\bigskip}

Isometric embeddings of the spacetime in five dimensions have been
considered with increasing interest in recent years. The development of the
braneworld scenario by Arkani-Hamed\cite{Arkani} {\it et al} and Randall \&
Sundrum\cite{Randall} has greatly favoured the idea among particle
physicists and cosmologists that in lower energy regime our Universe may
effectively be pictured as a four-dimensional hypersurface (brane)
isometrically embedded in a five-dimensional manifold (bulk). The study of
isometric embeddings has also been motivated by a proposal recently put
forward and developed by Wesson and others\cite{Wesson}, where vacuum $%
(4+1)- $dimensional field equations give rise to $(3+1)-$dimensional
equations with source, the energy-momentum tensor of which being related to
the extrinsic curvature of the spacetime \cite{maia}. Ultimately, all these
ideas seems to have drawn inspiration from the old Kaluza-Klein theory\cite
{kaluza}.

Embedding theories are naturally subject to embedding theorems of
differential geometry. Of interest here is a theorem due to Campbell and
Magaard\cite{Campbell}, which asserts that any n-dimensional Riemannian
analytic manifold can be locally and isometrically embedded in a ($n+1$%
)-dimensional Riemannian analytic manifold, where the Ricci tensor of the
latter vanishes. Campbell-Magaard's result may shed some light on our
understanding of the mathematical structure of embedding theories, and
applications of the theorem to the braneworld scenario, to five-dimensional
non-compactified Kaluza-Klein gravity as well as to superstring theory have
been found in recent years\cite{tavakol}. Extensions of the Campbell-Magaard
theorem to more general cases have also been obtained \cite{dahia1,dahia}.

Due to the high interest aroused by spacetime embedding theories, different
and specific types of embedding now start to be considered by theoretical
physicists. Such is the case of the so-called ideal embeddings, in which the
embedded space receives the ``least amount of tension'' from the surrounding
(pseudo-Euclidean) space \cite{stefan} {\em \ } An interesting kind of
embedding that may have a connection with particle physics is provided by
the so-called harmonic maps \cite{misner}. As models for physical theories,
harmonic maps lead to nonlinear field equations, which in many respects bear
a resemblance to Yang-Mills equations and to the Einstein equations for
gravitation. Harmonic maps are also linked to the non-linear sigma models,
first introduced by Schwinger, in the fifties, to describe massive, strongly
interacting particles \cite{Schwinger}. Recent developments in non-linear
sigma models include interaction with gravity, inflationary cosmological
models, Kaluza-Klein theories, among other topics\cite{Gibbons}.

Our aim here is to look at harmonic maps in the context of isometric
embeddings and show that they represent a non-empty subclass of the class of
all isometric embeddings when the target spaces are Ricci-flat with
codimension one \cite{Ulyanovsk}. The paper is organized as follows. In
section II we define the concept of isometric embedding of a n-dimensional
semi-Riemannian manifold in the class of ($n+1$)-dimensional Ricci-flat
semi-Riemannian manifolds. We then proceed to discuss the Campbell-Magaard
embedding theorem. In Section III we briefly define harmonic maps,\ discuss
how they are related to isometric embeddings and prove that for $n\geq 3$
any n-dimensional Lorentzian manifold can be isometrically and harmonically
embedded in the class of (n+1)-dimensional Ricci-flat semi-Riemannian
manifolds. In Section IV we give an example of a harmonic and isometric map
between two semi-Riemannian manifolds in which the target space is a
Ricci-flat manifold. An extension of these ideas to the case when the
embedding manifold is an Einstein space is briefly treated in Section V. In
Section VI we make some comments of geometric character on the meaning of
harmonicity for isometric embeddings. We end with a remark on the role
minimal embeddings could possibly play in the framework of embedding
theories of the spacetime.

\section{\protect\bigskip Isometric embeddings and the Campbell-Magaard
theorem}

The original version of the Campbell-Magaard theorem refers to Riemannian
manifolds, i.e. those endowed with positive definite metrics. It can be
shown, however, that this restrictive condition is not essential, so in what
follows we shall consider semi-Riemaniann manifolds with metrics of
indefinite signature. It is important to mention that throughout the paper
embeddings and metrics are supposed to be analytic.

Let ${\cal M}_{\pi }^{n+1}$ denote the class of all ($n+1$)-dimensional
manifolds $\overline{M}^{n+1}$which share some geometrical characteristic $%
\pi $ (For example, $\pi $ may express a restriction of the following kind:
to be Ricci-flat, to be an Einstein space, and so forth). A precise way of
defining isometric embeddings of a n-dimensional manifold $M^{n}$ in the
class of manifolds ${\cal M}_{\pi }^{n+1}$ is

\begin{definition}
We say that $M^{n}$ can be isometrically embedded in ${\cal M}_{\pi }^{n+1}$
if there exists at least one manifold $\overline{M}^{n+1}\in {\cal M}_{\pi
}^{n+1}$ in which $M^{n}$ can be isometrically embedded.
\end{definition}

Now, the Campbell-Magaard theorem states that any $M^{n}$ manifold can be
isometrically embedded in the class ${\cal M}_{\pi }^{n+1}$of vacuum
(Ricci-flat) spacetimes. It is known that this theorem can be demonstrated
by a method which consists in formulating the question as the initial value
problem in general relativity. Let us briefly review this technique.

Consider the metric of the ($n+1$)-dimensional space written in a Gaussian
coordinate system 
\begin{equation}
ds^{2}=\overline{g}_{ij}\left( x,\psi \right) dx^{i}dx^{j}+d\psi ^{2},
\label{hds2}
\end{equation}
where $x=\left( x^{1},...,x^{n}\right) $, and Latin indices run from $1$ to $%
n$ while Greek ones go from $1$ to $n+1.$

We can verify that the Einstein vacuum equations $R_{\mu \nu }=0$ expressed
in the above coordinates have the following structure\cite{Campbell}: 
\begin{eqnarray}
\frac{\partial ^{2}\overline{g}_{ij}}{\partial \psi ^{2}} &=&F_{ij}\left( 
\overline{g}_{lm},\frac{\partial \overline{g}_{lm}}{\partial \psi }\right) 
\label{dyn} \\
\nabla _{j}\left( K^{ij}-g^{ij}K\right)  &=&0  \label{c1} \\
R+K^{2}-K_{ij}K^{ij} &=&0,  \label{c2}
\end{eqnarray}
where $F_{ij}$ are analytical functions of their arguments, $\nabla _{j}$ is
the covariant derivative with respect to the induced metric $g_{ik}=%
\overline{g}_{ij}(x,\psi =const)$; $K=g^{ij}K_{ij}$, $R$ and\ $K_{ij}$\
denote, respectively, the scalar curvature and the extrinsic curvature of
the hypersurface $\psi =const.$ Recall that in the Gaussian coordinates of (%
\ref{hds2}) the extrinsic curvature assumes the simple form:
\begin{equation}
K_{ij}=-\frac{1}{2}\frac{\partial \overline{g}_{ij}}{\partial \psi }.
\label{ext}
\end{equation}
Owing to the Bianchi identities not all of the above set of equations are
independent. In fact, the second and third equations need to be imposed only
on some particular hypersurface, since they are propagated by the first one 
\cite{Campbell}. These equations will be referred to as{\it \ constraint
equations}.

To fix ideas let us chose, in the foliation defined above, a particular
hypersurface, say, $\psi =0$. According to the Cauchy-Kowalewskaya theorem 
\cite{Kowalewskaya}, equation (\ref{dyn}) always has a unique analytical
solution $\overline{g}_{ij}\left( x,\psi \right) $ provided that the
following analytical initial conditions are specified: 
\begin{eqnarray}
\overline{g}_{ij}\left( x,0\right) &=&g_{ij}\left( x\right)  \label{cig} \\
\left. \frac{\partial \overline{g}_{ij}}{\partial \psi }\right| _{\psi =0}
&=&-2K_{ij}\left( x\right) .  \label{cih}
\end{eqnarray}

From the perspective of the embedding problem these initial conditions
represent, respectively, the metric and the extrinsic curvature of the
hypersurface $\psi =0$, whereas the solution of equation (\ref{dyn}) gives
the metric of the $\left( n+1\right) -$dimensional space. Thus, if we can
guarantee that, for any \ given metric $g_{ik}(x)$, the constraint equations
always admit a solution, then the theorem is proved, since the solution
found $\overline{g}_{ij}\left( x,\psi \right) $ will satisfy $R_{\mu \nu }=0$%
. Clearly, the embedding map is then given by the equation $\psi =0$.

It turns out, as Magaard has proved\cite{Campbell}, that the constraint
equations always have a solution. Indeed, by simple counting operation we
can see that there are $n(n+1)/2$ unknown functions (the independent
elements of extrinsic curvature) and $n+1$ constraint equations, since the
metric $g_{ij}\left( x\right) $\ must be considered as a given datum. For $%
n\geq 3$, the number of variables is greater than the number of equations.
Thus using equation (\ref{c2}) to express one element of $K_{ij}$ in terms
of the others, Magaard has shown that equation (\ref{c1}) can be put in a
canonical form with respect to $n$ components of $K_{ij}$ conveniently
chosen. Then, once more, the Cauchy-Kowalewskaya theorem ensures the
existence of the solution. It is important to note that, as we have said
before, the number of variables is greater than the number of \ equations;
in this sense, we can say that there are $\left( n+1\right) \left(
n/2-1\right) $ degrees of freedom left over.

From this analysis we see that the embedding space is may not be unique.
Different choices of the free elements of $K_{ij}$ may lead to different
embedding spaces. Now let ${\cal S}_{\pi }^{n+1}$ be the set of all possible
embedding spaces for $M^{n}$ constructed by employing the Campbell-Magaard
method. Of course ${\cal S}_{\pi }^{n+1}\subset {\cal M}_{\pi }^{n+1}$. At
this point, we wonder if, among all the embeddings of $M^{n}$ in ${\cal S}%
_{\pi }^{n+1}$, there exists at least one that constitutes a harmonic map.
In the next section we define harmonic embeddings in ${\cal M}_{\pi }^{n+1}$.

\section{\protect\bigskip Harmonic maps}

\begin{definition}
We say that $M^{n}$ can be harmonically embedded in ${\cal M}_{\pi }^{n+1}$
if there exists at least one manifold $\overline{M}^{n+1}\in {\cal M}_{\pi
}^{n+1}$ in which $M^{n}$ can be harmonically embedded, i.e., there exists a
map $\phi :M^{n}\rightarrow \overline{M}^{n+1}$ that extremizes the action: 
\begin{equation}
I\left[ \phi \right] =\int d^{n}x\sqrt{\left| g\right| }g^{ij}\phi _{i}^{\mu
}\phi _{j}^{\nu }H_{\mu \nu }  \label{sigman}
\end{equation}
where $H_{\mu \nu }$ represents the metric tensor of $\overline{M}^{n+1}$
and \bigskip we are using the notation $\phi _{i}^{\mu }=\frac{\partial \phi
^{\mu }}{\partial x^{i}}.$
\end{definition}

Of course in order for $\phi $ to be also an isometric embedding in ${\cal M}%
_{\pi }^{n+1}$ we must have $\overline{M}^{n+1}\in {\cal S}_{\pi }^{n+1}$ (
here the property $\pi $ means that $H_{\mu \nu }$ is the metric of a vacuum
space). Assuming that $\overline{M}^{n+1}\in {\cal S}_{\pi }^{n+1}$, then,
as we know, there exist coordinates of $\overline{M}^{n+1}$ adapted to the
embedding, i.e., we can put the metric $H_{\mu \nu }$ in the Gaussian form 
\begin{equation}
ds^{2}=H_{ij}\left( x,\psi \right) dx^{i}dx^{j}+d\psi ^{2},
\end{equation}
with 
\begin{equation}
H_{ij}\left( x,0\right) =g_{ij}\left( x\right) .
\end{equation}
In these coordinates we have the following relations: 
\begin{eqnarray}
\phi ^{i} &=&x^{i}, \\
\phi ^{n+1} &=&\psi \left( x\right) .
\end{eqnarray}
Thus, the effective action becomes 
\begin{equation}
I\left[ \psi \right] =\int d^{n}x\sqrt{\left| g\right| }g^{ij}\left(
H_{ij}\left( x,\psi \right) +\psi _{i}\psi _{j}\right) .
\end{equation}
and the extremization condition 
\begin{equation}
\frac{\delta I}{\delta \psi }=0
\end{equation}
gives 
\begin{equation}
\frac{1}{\sqrt{\left| g\right| }}\partial _{i}\left( \sqrt{\left| g\right| }%
g^{ij}\psi _{j}\right) -\frac{1}{2}g^{ij}\partial _{\psi }H_{ij}\left(
x,\psi \right) =0
\end{equation}
From this equation we can conclude that the isometric embedding $\psi =0$ is
a harmonic map if and only if 
\begin{equation}
\frac{1}{2}g^{ij}\left. \partial _{\psi }H_{ij}\left( x,\psi \right) \right|
_{\psi =0}=0
\end{equation}
which, taking (\ref{ext}) into account, is equivalent to 
\begin{equation}
K=0.  \label{trK}
\end{equation}
Therefore the existence of the embedding space $\overline{M}^{n+1}$ depends
on the existence of a solution of the constraint equations (\ref{c1}), (\ref
{c2}) and (\ref{trK}). In fact, provided that these equations are satisfied
for some $K_{ij}$ and $g_{ij}$, the particular solution of (\ref{dyn}) with
these $K_{ij}$ and $g_{ij}$ as initial data, represents, in coordinates (\ref
{hds2}), a vacuum metric in $n+1$ dimensions. \ In this case, the map $\psi
=0$ will be an isometric and harmonic embedding. Thus, we have proved the
following:

A manifold $M^{n}$ can be isometrically and harmonically embedded in ${\cal M%
}_{\pi }^{n+1},$ if and only if there exist functions $K_{ij}$ which satisfy
the equations 
\[
\nabla _{j}\left( K^{ij}-g^{ij}K\right) =0 
\]
\[
R+K^{2}-K_{ij}K^{ij}=0, 
\]
\[
K=0. 
\]

\bigskip At this point it would be nice if we were able to prove that it is
always possible to choose the extrinsic curvature tensor in such a manner
that its components satisfy the constraint equations, i.e, that the
equations are not incompatible. It turns out that, this is the case, at
least for Lorentzian manifolds and $n\geqslant 3$, as it is shown in the
Appendix. Therefore, the following proposition holds true:

{\it Any analytical Lorentzian manifold }$M^{n}${\it \ }$\left( n\geq
3\right) ${\it \ can be analytically, isometrically and harmonically
embedded in the class of }$(n+1)-${\it \ dimensional vacuum spaces}.

\section{\protect\bigskip A simple example}

As a simple application of the ideas discussed above let us consider the
class of all $n$-dimensional manifolds whose scalar curvature $R$ vanishes.
Then, an obvious choice for the initial data required by the
Campbell-Magaard theorem is to take the extrisinc curvature tensor $K_{ij}=0$%
. \ In this case, the constraint equations are trivially satisfied.
Therefore, we conclude that a sufficient condition for a $n$-dimensional
manifold with vanishing curvature scalar $R$ to be isometrically and
harmonically embedded in a Ricci-flat ($n+1)$-dimensional manifold is that $%
K_{ij}=0$ . \bigskip This is the case, for instance, of the embedding of the 
$4-$dimensional Friedmann-Robertson-Walker metric 
\[
ds^{2}=dt^{2}-t(dx^{2}+dy^{2}+dz^{2}) 
\]
in the $5$-dimensional Ricci-flat space 
\[
ds^{2}=dt^{2}-t(dx^{2}+dy^{2}+dz^{2})-d\psi ^{2} 
\]
the embedding taking place for $\psi =0$.

This example may be easily generalized. Let 
\[
ds^{2}=g_{ij}(x)dx^{i}dx^{j}
\]
be the metric of $n$-dimensional space with $R=0$. It is then
straightforward to verify that this space can be isometrically and
harmonically embedded in the $(n+1)$-dimensional Ricci-flat space 
\[
ds^{2}=f(\psi )g_{ij}(x)dx^{i}dx^{j}-d\psi ^{2}
\]
where $f$ is a differentiable function with $f(0)=1$ and $f^{\prime }(0)=0$.
Indeed, in this case the extrinsic curvature tensor is given by 
\[
K_{ij}=\frac{-f^{\prime }(0)g_{ij}(x)}{2}=0
\]
hence all constraint equations are satisfied.\bigskip\ Since Ricci-flat
spaces have zero scalar curvature, we can state, as a  corollary of the
above, that any $n$-dimensional vacuum solution of general relativity may be
isometrically and harmonically embedded in a $(n+1)$-dimensional Ricci-flat
space{\it .}

\section{Harmonic maps and Einstein spaces\protect\bigskip}

All we have done up to this point may easily be carried over into more
general settings. Consider, for example, an extended version of the
Campbell-Magaard theorem \cite{dahia1}, which states that any $n$%
-dimensional semi-Riemannian manifold can be locally embedded in a $(n+1)$%
-dimensional \ Einstein space. In this case the Einstein equations $G_{\mu
\nu }=\Lambda g_{\mu \nu }$ are equivalent to the set 
\[
\frac{\partial ^{2}\overline{g}_{ij}}{\partial \psi ^{2}}=F_{ij}\left( 
\overline{g}_{lm},\frac{\partial g_{lm}}{\partial \psi },\Lambda \right) 
\]
\[
\nabla _{j}\left( K^{ij}-g^{ij}K\right) =0 
\]
\[
R+K^{2}-K_{ij}K^{ij}=-2\Lambda 
\]
where $\Lambda $ is a constant (usually referred to as the cosmological
constant of the embedding space). The extension of the \ previous theorem to
embeddings of n-dimensional ($n\geqslant 3$) Lorentzian manifolds in
Einstein spaces is readily seen to hold ( see Appendix ). Thus, we have the
following result:

{\it Let }$M_{\Lambda }^{n+1}${\it \ be the class of the }$(n+1)-${\it \
dimensional Einstein spaces with constant }$\Lambda ${\it . Then, any
analytical Lorentzian manifold }$M^{n}${\it \ }$\left( n\geq 3\right) ${\it %
\ can be analitically, isometrically and harmonically embedded in }$%
M_{\Lambda }^{n+1}${\it \ }.

\section{\protect\bigskip Final remarks}

In section III we have generically defined harmonic maps as those maps which
extremizes the action $I\left[ \phi \right] =\int d^{n}x\sqrt{\left|
g\right| }g^{ij}\phi _{i}^{\mu }\phi _{j}^{\nu }H_{\mu \nu }.$ Now if, in
addition, the embedding is assumed to be isometric, we have 
\[
\phi _{i}^{\mu }\phi _{j}^{\nu }H_{\mu \nu }=g_{ij} 
\]
In this case the action may be rewritten as 
\begin{equation}
I\left[ \phi \right] =\int d^{n}x\sqrt{\left| g\right| }g^{ij}g_{ij}=n\int
d^{n}x\sqrt{\left| g\right| }  \label{polyakov}
\end{equation}
Therefore we see that harmonic embeddings extremize the volume in the
induced metric $g_{ij}$. Let us make this statement more precise. If \ $%
M^{n} $ is isometrically and harmonically embedded in $\overline{M}^{n+1}$,
and if $D$ is a domain of $M^{n}$ with a regular boundary $\partial D$, then
the volume of $D$ in the induced metric \ is less ( greater, depending on
the signature of $g_{ij}$) than or equal to the volume of any other
submanifold of $\overline{M}^{n+1}$ with the same boundary. Embeddings which
satisfy this property are referred to, in the mathematical literature, as 
{\it minimal } \cite{Manfredo}. ( In fact, for Riemannian spaces it is known
that an isometric embedding is minimal if and only if it is a harmonic map 
\cite{Eeele}.) Let us just mention that an important characterization of a
minimally embedded space $M^{n}$\ lies in the fact that its mean curvature
vector $H$ vanishes for all points $p\in M^{n}$. In fact it is a well-known
fact that $H$ vanishes if and only if the trace of the extrinsic curvature $%
K_{ij}$ of the hypersurface $\psi =0$ is zero.

Finally, we would like to raise the question whether minimal embeddings may
play a role in higher-dimensional theories of the spacetime. In a paper
published some years ago Mc Manus\cite{Manus} posed a couple of questions
concerning the so-called induced matter models. In one of the questions it
was asked whether \ any energy-momentum tensor could be obtained by an
embedding mechanism. Due to the Campbell-Magaard theorem we now know that
the answer is yes, since any solution to the Einstein field equations
corresponding to any arbitrary energy-momentum tensor may be locally
embedded in a five-dimensional Ricci-flat space. The second question,
perhaps more relevant, was if there were a valid mechanism for determining
the choice of the four-dimensional hypersurfaces as the ones which would be
physically observables. We know that in the induced matter scheme the
dynamics of the surrounding space (bulk) is governed by the Einstein field
equations in vacuum. However, the choice of the foliations of the bulk by
four-dimensional hypersurfaces is entirely arbitrary. By requiring these
foliations to obey some (yet unknown) fundamental physical principle,
perhaps some progress could be made by linking the dynamics of the spacetime
to the dynamics of the bulk. In the specific case of minimal embeddings one
has a variational principle, namely, the extremization of the action defined
by the equation (\ref{polyakov}), i.e. the extremization of the volume of
the spacetime. This, at least, would put some constraints in the possible
choices of the foliation of the hypersurfaces. We hope these considerations
may stimulate work in this direction.

\section{\protect\bigskip Acknowledgements}

S.Chervon thanks the Physics Department of Universidade Federal da Paraiba
at Joao Pessoa (Brazil) for warm hospitality and CAPES for financial
support. C. Romero thanks CNPq for grant and financial support.

\section{\protect\bigskip \protect\bigskip Appendix}

Our aim in this appendix is to prove that, for $n\geq 3$, the constraint
equations of the section III can always be satisfied by an appropriate
choice of the extrinsic curvature tensor. Considering the possibility of
further applications to general relativity let us turn our attention to the
case when $M^{n}$ is Lorentzian.

For the sake of simplicity let us consider a Gaussian coordinate system of
the manifold $M^{n},$ where $x^{1}$ is a timelike coordinate. In these
coordinates the metric $g_{ij}$ assumes the following form 
\[
ds^{2}=-\left( dx^{1}\right) ^{2}+g_{AB}\left( x\right) dx^{A}dx^{B}, 
\]
where $A$ and $B$ run from $2$ to $n$.

Of course, $g_{AB}\left( x\right) $ is positive definite ( it is the induced
metric on the submanifold $x^{1}=const$), hence the diagonal components of $%
g_{AB}$ are all positive quantities. Thus, in terms of this coordinate
system the constraint equation (\ref{c1}) reads 
\begin{equation}
\frac{\partial K^{1i}}{\partial x^{1}}+\frac{\partial K^{Ai}}{\partial x^{A}}%
-\Gamma _{jm}^{i}K^{jm}-\Gamma _{jm}^{j}K^{im}=0,  \label{k1i}
\end{equation}
where $\Gamma _{jm}^{i}$ are the Christoffell symbols calculated with the
metric $g_{ij}$. As we emphasized previously for $n\geq 3$ there are more
components $K^{ij}$ than equations. On the other hand, since there are $n+2$
constraint equations we can choose $n+2$ components of $K^{ij}$ to be
treated as unknown functions in order to satisfy the constraint equations.
Let us select $K^{11},K^{12},...,K^{1n},K^{22}$ and \ $K^{33}.$ The
remaining components can be considered as independent functions, so they be
taken equal to zero.

After some algebra the constraints equations (\ref{c2}) and (\ref{trK}) may
be put in the following form: 
\begin{eqnarray}
-K^{11}+g_{22}K^{22}+g_{33}K^{33} &=&0  \label{k33} \\
\left( K^{11}\right) ^{2}-g_{AB}K^{1A}K^{1B}-\det g_{\left[ 2,3\right]
}K^{22}K^{33} &=&\frac{R}{2},  \label{k22}
\end{eqnarray}
where $\det g_{\left[ 2,3\right] }=g_{22}g_{33}-\left( g_{23}\right) ^{2}$
corresponds to the area measure of the surface spanned by the coordinates $%
x^{2}$ and $x^{3}$. Evidently $\det g_{\left[ 2,3\right] }>0,$ since a
positive definite metric satisfies the Cauchy-Schwarz inequality: $\Vert
X_{1}\Vert ^{2}\Vert X_{2}\Vert ^{2}\geq (X_{1},X_{2})^{2}$, where $X_{i}=%
\frac{\partial }{\partial x^{i}}$.

From (\ref{k33}) it follows that 
\begin{equation}
K^{33}=\frac{K^{11}-g_{22}K^{22}}{g_{33}}
\end{equation}
Substituting the above equation into (\ref{k22}) we find a quadratic
equation for $K^{22}:$%
\begin{equation}
\left( \frac{g_{22}}{g_{33}}\det g_{\left[ 2,3\right] }\right) \left(
K^{22}\right) ^{2}-\left( K^{11}\frac{\det g_{\left[ 2,3\right] }}{g_{33}}%
\right) K^{22}+\left( \left( K^{11}\right) ^{2}-g_{AB}K^{1A}K^{1B}-\frac{1}{2%
}R\right) =0.
\end{equation}
Now, this equation can be solved for $K^{22}$ if the discriminant is
non-negative. Clearly, in the case of a positive discriminant $K_{22}$ can
be written as an analytical function of $K^{1i}$ $\left( i=1,...,n\right) $
and the same holds for $K^{33}$ .

Surely, using the above relations we substitute the components $K^{22}$ and $%
K^{33}$ into (\ref{k1i}) obtain a first-order partial differential equation
for $K^{1i}$, which can be put in the following form 
\begin{equation}
\frac{\partial K^{1i}}{\partial x^{1}}=F^{i}\left( K^{1j},\frac{\partial
K^{1j}}{\partial x^{A}}\right) .
\end{equation}
Then, from the Cauchy-Kowalewskaya theorem, we conclude that if $F^{i}$ is
analytical in its arguments at some point, the origin, for example, then
there exists a unique analytical solution satisfying the initial condition 
\begin{equation}
K^{1i}\left( x^{1}=0,x^{2},...,x^{n}\right) =W^{i}\left(
x^{2},...,x^{n}\right) ,
\end{equation}
where $W^{i}$ are arbitrary analytical functions in the origin.

In order for $F^{i}$ to be analytic at the origin, it is sufficient that the
discriminant of the quadratic equation for $K^{22}$ be positive at this
point. In other words, we must have 
\begin{equation}
\left( W^{1}\frac{\det g_{\left[ 2,3\right] }}{g_{33}}\right) ^{2}-4\left( 
\frac{g_{22}}{g_{33}}\det g_{\left[ 2,3\right] }\right) \left( \left(
W^{1}\right) ^{2}-g_{AB}W^{A}W^{B}-\frac{1}{2}R\right) >0  \label{hold}
\end{equation}
It is not difficult to verify that it is always possible to have the above
inequality satisfied. Indeed, with the choice $W^{1}=0$ and $%
g_{AB}W^{A}W^{B}>-\frac{1}{2}R$, for instance, it is evident that (\ref{hold}%
) holds. Therefore, we conclude that any Lorentzian spacetime can be
minimally and analytically embedded in the class of five-dimensional vacuum
spaces.

This result can be trivially extended to the case of the embedding in the
class of Einstein spaces with a certain specified $\Lambda $. As far as the
constraint equations are concerned we have just to add the constant $\Lambda 
$ on the right hand side of equation (\ref{k22}). Then, following the same
procedure outlined above the existence of solutions of the constraint
equations is guaranteed \ just by requiring that $W^{1}=0$ and $%
g_{AB}W^{A}W^{B}>-\frac{1}{2}R-\Lambda .$

\end{document}